\documentclass[journal]{IEEEtran}
\usepackage[hidelinks]{hyperref}
\usepackage{graphicx}
\usepackage{algorithm}
\usepackage{algorithmic}
\usepackage[utf8]{inputenc}
\usepackage{babel}
\addto\captionsenglish{}
\usepackage{multirow,tabularx}
\usepackage{csquotes}
\usepackage{datetime}

\usepackage{bm}
\newdateformat{monthyear}{%
  \monthname[\THEMONTH], \THEYEAR}
\usepackage{amsfonts}
\usepackage[utf8]{inputenc}
\usepackage{wrapfig}
\usepackage{bookmark}
\DeclareMathSymbol{\shortminus}{\mathbin}{AMSa}{"39}
\usepackage{hyperref}
\usepackage{soul}
\usepackage[toc]{glossaries}
\usepackage[export]{adjustbox}
\usepackage{ragged2e}
\usepackage{array,varwidth,lipsum}
\usepackage{comment}
\usepackage{cases}
\usepackage{mathtools}
\usepackage{multirow}
\usepackage{multicol}
\usepackage{longtable}
\usepackage{amsmath,amssymb}
\usepackage{booktabs}
\usepackage[table,xcdraw]{xcolor}
\usepackage{cite}
\usepackage{skull}
\usepackage{lineno}
\usepackage{subfigure}

\begin{document}
\title{Online RMLSA in EONs with $A^3G$: Adaptive ACO with Augmentation of Graph}

\author{M Jyothi Kiran, Venkatesh Chebolu, Goutam Das,~\IEEEmembership{Member,~IEEE,}
        and~Raja Datta,~\IEEEmembership{Senior Member,~IEEE}}

\maketitle

\begin{abstract}
Routing and Spectrum Assignment (RSA) represents a significant challenge within Elastic Optical Networks (EONs), particularly in dynamic traffic scenarios where the network undergoes continuous changes. Integrating multiple modulation formats transforms it into Routing Modulation Level and Spectrum Assignment (RMLSA) problem, thereby making it more challenging. Traditionally, addressing the RSA problem involved identifying a fixed number of paths and subsequently allocating spectrum among them. Numerous heuristic and metaheuristic approaches have been proposed for RSA using this two-step methodology. However, solving for routing and assignment of spectrum independently is not recommended due to their interdependencies and their impact on resource utilization, fragmentation and bandwidth blocking probability. In this paper, we propose a novel approach to solve the RMLSA problem jointly in dynamic traffic scenarios, inspired by Ant Colony Optimization (ACO). This approach involves augmenting the network into an Auxiliary Graph and transforming conventional ACO into a constraint-based ACO variant that adapts to the constraints of EONs. This adaptation also includes an adaptive initiation process and an aggressive termination strategy aimed at achieving faster convergence. Moreover, we have introduced a novel objective/fitness function, to minimize average network fragmentation while ensuring optimal spectrum resource utilization, thereby reducing overall blocking probability.
\end{abstract}


\IEEEpeerreviewmaketitle

\section{Introduction}

\IEEEPARstart{G}{lobal} population having  access to the internet is growing exponentially. As per Cisco annual internet report 2021, there were 3.9 billion users (51\% of the global population) and it was predicted to increase to around 5.3 billion (almost 61\% of the global population) by 2023. Moreover, internet traffic in recent times is also rapidly increasing in the backbone network because of the emerging applications such as high definition televisions, real-time video conferencing, augmented reality\cite{growth}.  Conventional optical networks cannot support such higher user base and data rates because of it's fixed grid design and associated inflexibility. Elastic Optical Networks (EONs) emerged as a potential solution to solve the above mentioned problem as it can efficiently utilize the c-band optical spectrum \cite{intro1,intro2}. Routing and Wavelength Assignment (RWA) \cite{ozdaglar2003routing,zang2000review,rouskas2001routing,ramaswami1994optimal} is a significant networking problem for Wavelength Division Multiplexing (WDM) in conventional optical networks. RWA algorithms provide requested connection by allocating appropriate lightpaths in the network. As conventional WDM optical networks move towards EONs, a new networking problem of Routing and Spectrum Assignment (RSA) evolved, which is analogous to the RWA in WDM networks\cite{intro2}. The RSA problem, when incorporating multiple modulation formats, evolves into the Routing, Modulation Level and Spectrum Assignment (RMLSA) problem.

\subsection{Related works}
In this section, works related to RSA/RMLSA are presented, categorized into three parts corresponding to ILPs, heuristics, and metaheuristics.
\subsubsection{\textbf{Approaches based on ILP}}
 The works presented in \cite{static_ILP1_capucho2013ilp} and \cite{static_ILP2_klinkowski2011routing} introduce an Integer Linear Programming (ILP) model designed to solve the Static RSA problem, primarily aiming to minimize the total network frequency slots using a single modulation format per connection. In contrast, the methodology outlined in \cite{static_ILP3_christodoulopoulos2011elastic} employs multiple modulation formats. Moreover, the authors in \cite{dynamic_ILP1_zhang2023optimal,dynamic_ILP2_chen2015efficient,dynamic_ILP3_ruan2014dynamic} discuss ILP formulations for dynamic traffic. Specifically, \cite{dynamic_ILP1_zhang2023optimal} focuses on minimizing the number of hops using single modulation formats, while \cite{dynamic_ILP2_chen2015efficient} considers both the minimization of hops and overlooks fragmentation concerns.

\subsubsection{\textbf{Heuristics solving RSA}}
Heuristic algorithms have become prominent in solving the RSA problem due to their ability to provide solutions with lower complexity compared to ILP. Their significance is further underscored in dynamic traffic scenarios, where rapid solution generation is essential. Heuristics were employed to solve the RSA problem for dynamic traffic scenarios in \cite{dynamic_hueristic1_wan2011dynamic,dynamic_hueristic2_takagi2011dynamic,dynamic_hueristic3_costa2021dynamic,dynamic_hueristic6_adhikari2020impact,dynamic_hueristic4_khan2019online,dynamic_hueristic5_wang2015distance,dynamic_hueristic7_chebolu2022robust}. Specifically, the authors in \cite{dynamic_hueristic1_wan2011dynamic} proposed three heuristics to address dynamic RMLSA while the authors in \cite{dynamic_hueristic2_takagi2011dynamic} and \cite{dynamic_hueristic3_costa2021dynamic} developed distance-adaptive heuristics aimed at reducing the bandwidth blocking probability without addressing fragmentation. In \cite{dynamic_hueristic6_adhikari2020impact} and \cite{dynamic_hueristic4_khan2019online}, the dynamic RSA and RMLSA problems were respectively solved considering the network fragmentation. In \cite{dynamic_hueristic5_wang2015distance} and \cite{dynamic_hueristic7_chebolu2022robust} the authors used heuristics for dynamic traffic RSA that also considered network survivability. 

\subsubsection{\textbf{Emergence of metaheuristics}}
Researchers have investigated metaheuristics to address the RSA problem in EONs. Metaheuristics are strategies designed to guide the search process, aiming to efficiently explore the search space. These strategies range from simple local search procedures to complex learning processes. The authors in \cite{meta_ABC_xu2018routing} proposed an artificial bee colony optimization to solve the routing problem, while in \cite{meta_genetic_dinarte2021routing}, a genetic algorithm was proposed to determine whether to perform routing or spectrum assignment first. In \cite{meta_hynea_kumar2023traffic}, the spotted hyena optimizer was utilized in EONs for both routing and rerouting. The metaheuristic inspired by the natural behavior of ants, known as Ant Colony Optimization (ACO), was first reported in \cite{ants_dorigo1996ant}. It was later adapted for use in communication networks for routing purposes as the AntNet algorithm, as noted in \cite{ants_in_comm_di1998antnet}. ACO has also been applied in optical networks to address routing problems, such as in \cite{ants_rwa_pavani2006traffic} and \cite{ants_rsa_de2021provisioning}, where it was used for rerouting via the crankback mechanism corresponding to RWA and RSA respectively. In \cite{comp_nan2022routing}, the routing in RSA problem was solved using ACO taking fragmentation into account.
\subsection{Gap Analysis}
Here the issues that were overlooked earlier, have been listed and analyzed.
\begin{itemize}
    \item In the previous subsection we find that, in general, the RSA problem has been bifurcated into two distinct tasks, i.e., (a) routing and (b) assignment of spectrum. Further, it has been highlighted in \cite{meta_genetic_dinarte2021routing} that tackling the RSA problem jointly is exceedingly complex. However, we feel that a joint approach for optimal resource allocation is necessary due to the strong interdependence of routing and assignment of spectrum. 
    \item In addressing the problem of RSA, particularly in the context of dynamic traffic, the primary factors influencing network performance are spectrum utilization and network fragmentation. The network fragmentation problem is found to be a crucial consideration with the dynamic traffic, because the connections arrive and depart at varying intervals creating fragments in one or multiple links of the network. The standard approach to spectrum allocation involves allocating the required spectrum from the lowest indexed slot among a fixed number of established paths. However, this approach may not be advisable in dynamic traffic scenarios, as fragments can occur at various positions within the range of Frequency Slot Units (FSUs) at different times. 
\end{itemize}

\subsection{Approach to address the gaps}

We previously stated that the routing and assignment of spectrum are interdependent and preferably be solved jointly for optimal resource allocations. To address this, we have developed a metaheuristic algorithm inspired by ACO. By transforming the conventional ACO into a constraint-based ACO, we incorporate all necessary constraints to reduce complexity, enabling the ants in the ACO to effectively and jointly search for solutions. As we tackle dynamic traffic, where network conditions change rapidly, our algorithm is designed to adapt to these changes and deliver prompt solutions. This adaptation to deliver quicker solutions is implemented in three phases: augmentation of the graph, determination of the number of ants to be deployed for solution search, and aggressive termination. Further, as fragmentation is crucial in dynamic connections, we have developed an objective function, referred to as the fitness function in metaheuristic algorithms that aims to allocate spectrum within the available fragments. The proposed fitness function also ensures minimal use of resources so that resource overutilization may be avoided, especially in dynamic scenarios where the future connections are unknown. To the best of our knowledge, the problem of RMLSA for dynamic traffic has not been jointly addressed, especially with a focus on both spectrum utilization and network fragmentation within the domain of EONs.

\subsection{Contributions}
\begin{itemize}
    \item[\textbf{1.}] First, we generate an auxiliary graph by augmenting the EON with available inputs and related constraints. 
    \item[\textbf{2.}] Next, we propose a constraint based metaheuristic algorithm, inspired by ACO, for solving RMLSA on the auxiliary graph. 
    \item[\textbf{3.}] The proposed algorithm has adaptive initialization and aggressive termination mechanisms that facilitates accelerated convergence.
    \item[\textbf{4.}] Moreover, we have developed a fitness function for the algorithm that strategically utilizes the spectrum effectively and improves average network fragmentation.
    \item[\textbf{5.}]  We have conducted extensive simulations to demonstrate that our proposal achieves a $5.4\%$ reduction in network resource usage and a $14\%$ decrease in network average fragmentation. This lower resource utilization and reduced average network fragmentation enable the network to handle an additional 1.1 Tbps of load and a lesser bandwidth blocking probability of $13\%$ at lower arrival rates and $34\%$ at higher arrival rates compared to the existing ones.
\end{itemize}

\section{System Model}
In this section, we present the network model and delineate the inputs utilized for the formulation and solution of the proposed metaheuristic.

\subsection{Network Model}
We model the network as $G(V, E, FSU)$ where $v_i \in V$ is a vertex and $e_{i,j} \in E$ is the edge between nodes $i,j$. We define $fsu_{i,j,k} \in FSU$ as $k^{th}$ FSU index in link $(i,j)$. The value of $fsu_{i,j,k}$ can be either  ‘0’ or ‘1’ stating whether it is un-occupied or occupied respectively. $d_{i,j}$ is the distance corresponding to edge $e_{i,j}$. We consider $N$ as the total number of FSUs in a link. $M$ is the set of modulation formats allowed in the allocation of the spectrum and $m_l \in M$ corresponds to each modulation format. The maximum transmission reach for a particular modulation format is denoted as set $D$, which contains $\left \{d_{m_1},d_{m_2},...,d_{m_M} \right \}$ corresponding to $\left \{m_1,m_2,...,m_M \right \}$ respectively.
\subsection{Inputs}
The inputs provided are source, $s \in V$, destination, $d \in V$ and a data rate, $\rho$ Gbps required for a connection. Additionally, the current state of the network is provided at the requested instance. The data rate is then converted to a number of FSUs that are required to allocate the connection for different modulation formats. The minimum slots required to satisfy the requested demand is given by $FS_{m_{1}} = \left \lceil \frac{\rho}{10*1}  \right \rceil, FS_{m_{2}} = \left \lceil \frac{\rho}{10*2}  \right \rceil,..., FS_{m_{M}} = \left \lceil \frac{\rho}{10*M}  \right \rceil$ respectively for $m_1, m_2,..., m_M$.

With the mentioned system model and the provided inputs, we propose a metaheuristic algorithm for solving the RSA problem. In the next two sections, we define our fitness function and our novel graph augmentation method to implement the constraints of RSA within the ACO framework.

\section{Fitness function and its Significance}

In this section, we present the formulation of the fitness function of our metaheuristic algorithm and discuss the underlying reasons for its selection.

\subsection{The Fitness Function}
The formulation of the fitness function is the most crucial aspect of a metaheuristic algorithm as it is primarily responsible for identifying optimal solutions. Effective fragmentation along with minimal spectrum utilization is a key parameter that influences the allocation of the maximum number of connections. By considering these two parameters, we developed a fitness function and as follows:

\begin{equation}
    \text{Fitness Function} = \frac{\Delta F}{2LT} + FS_m * LT
    \label{FitnessFunction}
\end{equation}

\noindent where $\Delta F = $ Change of fragments in the traversed path,\\
 $LT = $ Length of the traversed path.

\subsection{Significance of the fitness function}
In the fitness function described in (\ref{FitnessFunction}), it is evident that two distinct components exist: one corresponding to fragmentation and the other to resource utilization. This deliberate choice was made to optimize spectrum utilization while concurrently addressing network fragmentation. A comprehensive explanation of this strategic decision will be provided in the forthcoming subsections.

 \textit{Significance of fragmentation:} It is a well established fact that the assignment of spectrum slots at any index of a link gives rise to one of the three distinct outcomes: (a) generation of an additional fragment, (b) reduction of a fragment by accommodating within an existing fragment, or (c) the preservation of a fragment without any modification. Thus, the alterations in fragments within a link can only manifest as [1, -1, 0], signifying the three aforementioned scenarios respectively, with no other potential values emerging. 

\begin{figure}[h]
    \centering
    \includegraphics[width=0.9\columnwidth,keepaspectratio]{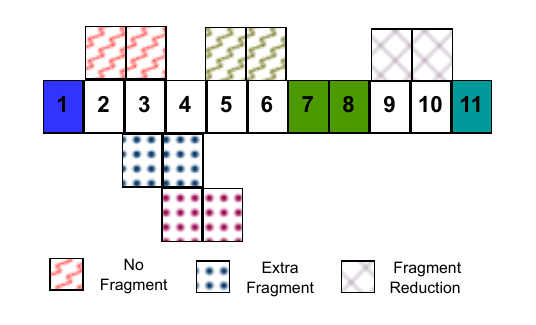}
    \caption{Fragment Illustration}
    \label{Fitness_Fragments}
\end{figure}

To illustrate that, we considered a link as an example with few FSUs already filled (solid boxes) is shown in Fig. \ref{Fitness_Fragments}. We find that two FSUs are required for the allocation, where the allocation at FSU with indices 2 or 5 does not create fragments, whereas, allocating FSU with indices 3 or 4 creates an additional fragment. Conversely, allocating FSU with index 9 results in the reduction of the existing fragment. No other scenarios within this link is possible.

We see in (\ref{FitnessFunction}) that the number of links traversed in the path is LT and so the cumulative change in fragments in the entire path fall in the range [-LT, LT]. Therefore, normalizing this range by twice the total number of links traversed, converges it to [-1/2, 1/2].

 \textit{Significance of resource utilization:} 
 The second component signifies the total number of FSUs needed to facilitate the allocation of a requested connection. Further, the required number of FSUs within a link may vary from 1 to N based on the requested data rate, thereby setting the range to be [1, NLT] for LT number of links traversed.

 \textit{Merit of integrating fragmentation and resource utilization:} We find that while choosing a solution from the available options based on the fitness value, the solution with the minimum fitness value will correspond to the least spectrum utilization. This is because the lower limit of the second component is higher than the upper limit of the first component. Given that multiple solutions may exist that takes same number of FSUs, the first component becomes significant. And so, it chooses the solution with the lowest value of the first component that directly reflects the least change of fragments. Moreover, prioritizing spectrum utilization over fragmentation is essential because the dynamic traffic that necessitates efficient spectrum usage due to unknown future connections.

\section{Augmentation of Graph}
 ACO algorithms are widely recognized for its application in routing problems across various domains. This includes EONs, where it identifies routes between the node pairs based on the pheromone values. However, it is essential to solve both the routing and assignment of spectrum jointly for the solutions to be optimal. To achieve this, we first augment the network to provide the ants with crucial information. This involves constructing an auxiliary graph from the original network, providing all necessary information for problem-solving. By creating this auxiliary graph, the ants are directed towards optimal solutions, avoiding suboptimal paths. This modification not only helps in solving RMSLA problem jointly but also make our proposed algorithm fully dynamic and adaptable to network changes. 

\subsection{Generation of a Auxiliary Graph}

\begin{figure}[h]
    \centering
    \includegraphics[width=0.9\columnwidth,keepaspectratio]{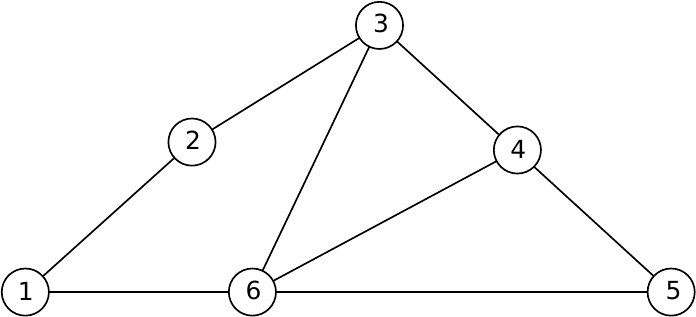}
    \caption{6 Node Network}
    \label{Six_node_Network}
\end{figure}

To facilitate the ants exploration in finding a solution, it is crucial to provide them with links that include the appropriate pheromone levels. However, the actual network comprises of links that correspond to the distances between the nodes. For the ants to effectively and jointly determine a solution, they need a comprehensive set of links representing every possible solution. To achieve this, we develop an auxiliary graph that contains a comprehensive set of virtual links corresponding to every possible FSU index and modulation format at each node. These virtual links, also referred to as auxiliary links, enable the ants to gather information on each potential conditions such as various routes, modulation formats and FSU indices. By populating these virtual links, we ensure that the ants have the flexibility to explore all possible conditions and obtain the necessary information to make informed decisions. An example network with $6$ nodes, $8$ links and $N$ FSUs is shown in Fig. \ref{Six_node_Network}.

\begin{figure}[h]
    \centering
    \includegraphics[width=0.9\columnwidth,keepaspectratio]{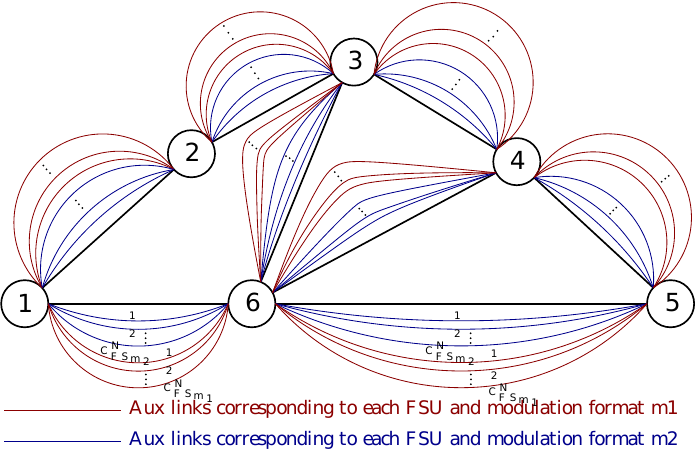}
    \caption{Auxiliary Graph}
    \label{Aux_Graph_prelim}
\end{figure}

Considering two modulation formats $m_1$ and $m_2$, here we construct an auxiliary graph that is based on the network shown in Fig. \ref{Six_node_Network}. We find that the number of slots required are $FS_{m_1}$ and $FS_{m_2}$ for $m_1$ and $m_2$ respectively. Each auxiliary link corresponds to one of the possible ways to determine the FSUs in a link to allocate the connection for spectrum allocation. It can be observed that the number of combinations that can be formed in a link is $C_{FS_m}^N$ for each modulation format, as we need to select $FS_m$ slots from $N$ slots in a link. Therefore, the total number of auxiliary links for each link turns out to be $C_{FS_{m_1}}^N+ C_{FS_{m_2}}^N$. Thus, making the total auxiliary links for $L$ links to be $(C_{FS_{m_1}}^N+ C_{FS_{m_2}}^N) \times L$ as shown in Fig. \ref{Aux_Graph_prelim}. In general, for $M$ modulation formats, the total number of links are $(C_{FS_{m_1}}^N+ C_{FS_{m_2}}^N+...+ C_{FS_{m_M}}^N) \times L$. 

The total number of auxiliary links here is important as it helps to determine the overall size and complexity of the graph, which can impact the efficiency of algorithm and computations performed on it. If the value of the total number of auxiliary links turns out to be very large, it can become computationally heavy, making it ineffective for real-time networks. This can significantly impact the performance of the network, leading to delays and inefficiencies in processing data.

To address this issue, we constrained the exploration region while ensuring all potential solutions are still considered. This was accomplished by transforming the conventional ACO algorithm into a constraint-based ACO. In this modified approach, we utilized the available data, which includes (a) source-destination pair, (b) the requested data rate between them and (c) the state of the network at the requested instance. This information is used to limit the exploration region and enforce necessary constraints for solving the RSA problem. The reduction associated with each parameter is detailed in the subsequent subsections.

\subsubsection{Reduction due to continuity}

Given the necessity of maintaining continuity constraints in transparent optical networks, it is mandatory that no changes occur in the FSU index at intermediate nodes. This implies that the FSU indices and modulation format determined at the source node must remain constant until the destination node is reached. Therefore, it is sufficient to verify the availability of the selected FSU indices at the intermediate nodes. To comply with this constraint, we propose deleting all virtual links in the network and preserving only the auxiliary links at the source node. For instance, by designating node $1$ as the source, the resultant auxiliary graph will contain solely the auxiliary links at the source node, with all other virtual links eliminated and shown in Fig. \ref{Aux_after_cg_cn}.

As a result of this constraint, the number of auxiliary links present in the auxiliary graph gets reduced to  $ (C_{FS_{m_1}}^N+ C_{FS_{m_2}}^N+...+ C_{FS_{m_M}}^N) \times deg(s) + (L - deg(s)) $, where $deg(s)$ represents the degree of the source node. This means that the total number of auxiliary links present in the network is determined solely by the degree of the source node.

 In a standard 14 node NSFNET, the percentage reduction in the total number of links is calculated and listed below.

 \begin{table}[h]
\centering
\begin{tabular}{|c|c|}
\hline
\textbf{Degree}                          & \textbf{Reduction in percentage of links} \\ \hline
Maximum Degree = 4                       & 80.91 \%                                  \\ \hline
\multicolumn{1}{|l|}{Minimum Degree = 2} & 90.47\%                                   \\ \hline
\end{tabular}
\end{table}

\subsubsection{Reduction due to Contiguity}
The contiguity constraint is vital while solving the RSA problem. This constraint mandates that the slots required to satisfy a demand must be contiguous. By applying this constraint, we have effectively reduced the number of auxiliary links in the auxiliary graph still not excluding any potential solutions.

As detailed and shown in Fig. \ref{Aux_Graph_prelim}, each modulation format associated with a link has $C^N_{FS_m}$ auxiliary links. This count results from choosing $FS_m$ slots from a total of $N $ slots. However, the contiguity constraint requires that the slots be contiguous. To satisfy this constraint while keeping the $FS_m$ slots contiguous, the number of potential combinations is reduced to a maximum of $N-FS_m +1$, thereby significantly decreasing the number of auxiliary links.

An instance is considered where 1 slot is required for $m_1$ and 2 slots are required for $m_2$ for a successful allocation. The number of combinations formed without the contiguity constraint and with contiguity constraint is calculated and listed for various values of $N$. Additionally, the percentage reduction in these combinations is also also listed for the same values of $N$. 

\begin{table}[h]
\centering
\begin{tabular}{|c|c|c|c|}
\hline
\textbf{FSUs}                          & \textbf{without Contiguity}  & \textbf{with Contiguity}   & \textbf{Reduction}\\ \hline
N = 6 & $C_2^6 + C_1^6$                       & $6+(6-1)$  & 53\%                                 \\ \hline
N = 100  & $C_2^{100} + C_1^{100}$                       & $100+(100-1)$  & 96 \%                                   \\ \hline

N = 320  & $C_2^{320} + C_1^{320}$                       & $320+(320-1)$  & 98.75 \%                                   \\ \hline
\end{tabular}
\end{table}

\begin{figure}[h]
    \centering
    \includegraphics[width=0.9\columnwidth,keepaspectratio]{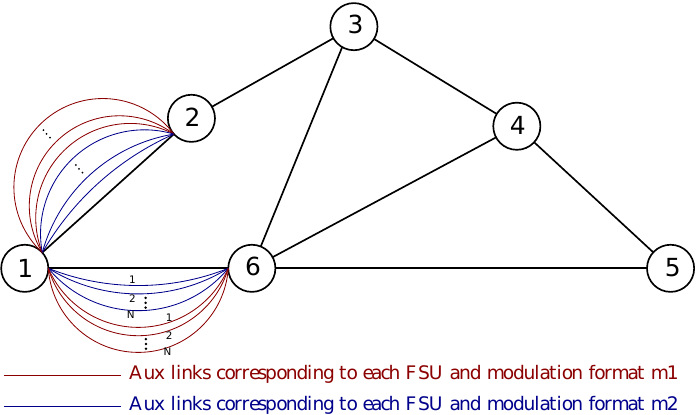}
    \caption{Auxiliary Graph after reduction due to continuity and contiguity}
    \label{Aux_after_cg_cn}
\end{figure}

  Thus, all feasible auxiliary links at the source node, with each link corresponding to the first available FSU index and different modulation formats that can facilitate the connection are illustrated in Fig. \ref{Aux_after_cg_cn}. This reduction, when combined with the non-overlapping constraint and the current state of the network, becomes even more advantageous. A detailed explanation of these benefits is provided in the subsequent subsection.

\subsubsection{Reduction due to state of the Network}

As mentioned earlier, we are solving for dynamic traffic, where the network state rapidly evolves due to the continuous arrival and departure of requests. We have utilized this rapidly changing complexity to our advantage. To avoid overlapping, unused auxiliary links are eliminated once allocated to other connections. In addition, by integrating the contiguity constraint, we further eliminate auxiliary links that are not potential solutions. This reduction in auxiliary links directly impacts the exploration opportunities available to the ants, potentially leading to quicker convergence of solutions.

\begin{figure}[h]
    \centering
    \includegraphics[width=0.9\columnwidth,keepaspectratio]{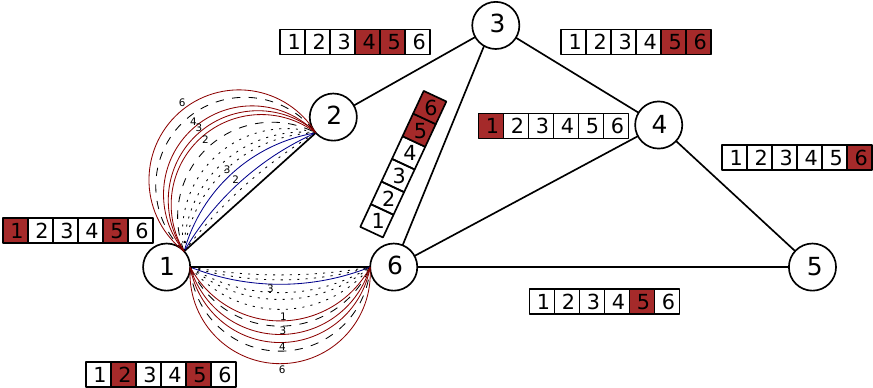}
    \caption{Auxiliary Graph}
    \label{Final_Aux_graph}
\end{figure}

We assume that the same number of FSUs are required as was mentioned previously in the auxiliary graph that has been developed. We further consider that 6 slots are present in each link and a few slots among them were already occupied as indicated by solid boxes in Fig. \ref{Final_Aux_graph}. Thus, the final auxiliary graph is updated by considering all the three constraints namely the contiguity constraint, non-overlapping constraint and the current state of the network, as shown in Fig. \ref{Final_Aux_graph}. We observe that with 4 slots occupied from the links at source node, the number of auxiliary links is further reduced by 54\%.

It is evident that the auxiliary graph produced for a specific source-destination pair at a particular instance is entirely dynamic, making it adaptive to the changes in the network. We have outlined the procedure in Algorithm 1 for generating an auxiliary graph upon the arrival of a request.

\begin{algorithm}[h]
\caption{Procedure for Auxiliary Graph Generation}
\begin{algorithmic}[1]
 \renewcommand{\algorithmicrequire}{\textbf{Input:}}
    \renewcommand{\algorithmicensure}{\textbf{Output:}}
    \REQUIRE $G(V,E,FSU), source(s), FS, M$
    \ENSURE $AG(V,E,AE)$
    \STATE $AG(V, E, AE) =$ empty graph with vertices V, edges E, and auxiliary-edge mappings AE
    \FOR{$i=1$ to $V$}
        \FOR{$j=1$ to $V$}
            \STATE $e_{i,j}= e_{i,j}$
        \ENDFOR
    \ENDFOR
    \FOR{$i$ in $\mathcal{N}(s)$}
        \FOR{$l=1$ to $M$}
            \STATE $temp =$ array of zeros with length $FS_{m_l}$
            \FOR{$k=1$ to $N$}
                \STATE $test =$ compare($temp,fsu_{s,i,[k:k+FS_{m_l}-1]}$)
                \IF{$test = true$}
                    \STATE $ae_{s,i,m_l}^k = 1$ 
                \ELSE
                    \STATE $ae_{s,i,m_l}^k = 0$
                \ENDIF
            \ENDFOR
        \ENDFOR
    \ENDFOR
\end{algorithmic}
\end{algorithm}

So, we define the auxiliary graph $AG(V,E,AE)$ as a graph derived from $G(V, E, FSU)$ where $v_i \in V$ is the vertex and $e_{i,j} \in E$ is the edge between nodes $i,j$, which is similar to the original graph. $ae_{s,i,m_l}^k$ is an auxiliary link generated at source node to all its connected neighbours $\mathcal{N}(s)$ for a modulation format $m_l \in M$ corresponding to FSU index $k$. The auxiliary links generated are the first slots that are feasible for provisioning a connection.

\section{Adaptive ACO on Auxiliary Graph : $A^3G$}

In this section, a metaheuristic optimization algorithm that is inspired from ACO has been developed on the auxiliary graph to optimize the fitness function. 

\subsection{Initial pheromone concentration}
To begin the optimization process, we fed initial pheromone concentrations into all the links in the auxiliary graph. This initial pheromone concentration serves as a basis for the meta-heuristic algorithm to begin exploring possible solutions and identifying the optimal path. The initial pheromone deposition is shown in equations (\ref{Initial_Pheromone_edge}) and (\ref{Initial_Pheromone_AuxEdge}), where $\tau_{i,j}$ indicates pheromone concentration of edge $e_{i,j}$ and $ \tau_{s,j,m_l}^k$ corresponding to the pheromone concentration in the auxiliary edge $ae_{s,j,m_l}^k$.

\begin{equation}
    \tau_{i,j} = \frac{1}{d_{i,j}} ;  ~~ \forall ~ i \in V-\left \{ s \right \} ~ ,  ~ \forall~ j \in V-\left \{ s \right \}
    \label{Initial_Pheromone_edge}
\end{equation}

\begin{equation}
    \tau_{s,j,m_l}^k = \frac{1}{(m_l + k)}; ~ \forall j  \in \mathcal{N}(s) ,~ \forall k \in N,~ \forall m_l \in M 
    \label{Initial_Pheromone_AuxEdge}
\end{equation}

\subsection{Initialization of Ants}

After the initial pheromone concentration is distributed across all links in the auxiliary graph, a total of A ants are loaded at the source node to explore the network. The number of ants is chosen according to (\ref{Ants_Initialization}).

\begin{equation}
 A = z \times degree(source(AG))
 \label{Ants_Initialization}
\end{equation}

\noindent where $z$ is a scaling parameter. From (\ref{Ants_Initialization}) we observe that the total number of ants will depend on the degree of the auxiliary graph. As demonstrated in the previous section, the auxiliary graph is influenced by the degree of the source and the state of the network at the requested instant. Therefore, the initialization of the total number of ants, which impacts the complexity, is made adaptive to accelerate the convergence of the algorithm. In this study, the value of $z$ is considered to be between 0.5 to 5, and the results are presented in the results section.

\subsection{Exploration and Exploitation}
Exploration and exploitation are fundamental features of metaheuristic algorithms. Exploration involves actions that enable the agent to uncover new aspects of the environment, while exploitation involves utilizing previously acquired knowledge. An agent that solely exploits past experiences risks of becoming trapped in a suboptimal policy. Conversely, an agent that only explores may never converge on an optimal policy. Therefore, there must be a balance between exploration and exploitation to identify the optimal policy that maximizes rewards.

Our developed metaheuristic algorithm incorporates an adaptive allocation of ants for exploration and exploitation. During the initial iteration, all ants are allocated to exploration due to the absence of prior knowledge about the network. For the second iteration, the information obtained from the first iteration is used for exploitation. To avoid suboptimal solutions, 50\% of the ants are allocated for exploration and the remaining 50\% for exploitation. This exploration and exploitation continues in subsequent iterations, with the allocation of specific number of ants as defined by (\ref{Exploration_Number}) and (\ref{Exploitation_Number}) respectively.

\begin{equation}
    \textit{Number of ants for exploration} = \left \lceil  \frac{A}{Ite} \right \rceil 
    \label{Exploration_Number}
\end{equation}

\begin{equation}
    \textit{Number of ants for exploitation} = A - \left \lceil  \frac{A}{Ite} \right \rceil 
    \label{Exploitation_Number}
\end{equation}

To address this mathematically, ants are labeled from $1$ to $A$. During the iteration $Ite$, ants labeled from 1 to $\left \lceil  \frac{A}{Ite} \right \rceil$ are designated for exploration, and the rest for exploitation. This labeling method will not impact either exploration or exploitation, as each ant operates independently within a given iteration, without influencing the other ants. A comprehensive explanation of exploration and exploitation is provided in the subsequent subsections.  

\begin{algorithm}
\caption{Ant Colony Optimization for Path Finding in SDM EONs}
\textbf{Input:} Source, destination, graph \\
\textbf{Output:} Colony of ants with distance and fitness values

\begin{algorithmic}[1]
\FOR{all iterations}
    \FOR{each ant}
        \STATE Start from the source node
        \STATE Choose a link with probability according to Equation (2)
        \STATE Update the distance and selected core
        \WHILE{destination is not reached and no impasse is encountered}
            \STATE Fix core and find the next node based on probability in Equation (3)
        \ENDWHILE
            \STATE Calculate fitness of the ant using Equation (4)
    \ENDFOR
    \STATE Update pheromones on paths using Equation (5)
\ENDFOR

\end{algorithmic}
\end{algorithm}

\subsection{Exploration} 
 Due to their inherent nature, the ants follow paths with higher pheromone concentrations. To model this behavior, we calculate a probabilistic value based on the pheromone concentrations on all possible links in the auxiliary graph. The probability of selecting an edge $e_{i,j}$ is represented as $Pe_{i,j}$, which signifies the probability of choosing node $j$ when the ant is at node $i$, is given in (\ref{Probability_Equation_Edge}. Additionally, the probability of selecting an auxiliary edge $ae_{s,j,m_l}^k$ is denoted as $Pae_{s,j,m_l}^k$. This denotes the probability of choosing an auxiliary link from source $s$ to node $j$, with FSU index $k$ and modulation format $m_l$ as given in (\ref{Probability_Equation_Aux_Edge}). Based on the generated link probabilities, a path is formed that satisfies the given constraints, thus transforming the conventional ACO algorithm into a constraint-based ACO. This adaptation involves maintaining specific conditions at each node when selecting the subsequent link. The procedure for ant exploration is given in Algorithm 2.

\begin{equation}
Pe_{i,j} = \frac{\tau _{i,j} }{\sum_{j \in V-\left \{ s \right \}}^{}\tau _{i,j}}
    \label{Probability_Equation_Edge}
\end{equation}

\begin{equation}
Pae_{s,j,m_l}^{k} = \frac{\tau_{s,j,m_l}^k}{\sum_{j \in \mathcal{N}(s) }\sum_{k \in N} \sum_{m_l \in M} \tau_{s,j,m_l}^k} 
    \label{Probability_Equation_Aux_Edge}
\end{equation}

\begin{algorithm}[h!]
\caption{Procedure for exploration/explotation by Ants}
\begin{algorithmic}[1]
\FOR{$a=1$ to $A$ (all Ants)}
    \STATE Initialize ant, $a$: starting at node $s$ with dis $0$
    \FOR{$j=2$ to $V$} 
        \STATE Set currentNode ($cN$) as  last visited node by ant $a$
        \IF{$j=2$} 
       \STATE \COMMENT{Searching for first intermediate node}
            \IF{$ deg(AG(s)) ==0$}
                \STATE  Exit   \textcolor{teal}{No Exploration Possible}
            \ELSE
                \STATE Pick  nextNode $nN$ using roulette wheel based on probabilities calculated in (\ref{Probability_Equation_Aux_Edge}) 
                \STATE Update info:$ant[a].fsuindex,mod \gets k,m$, $nextNode \gets nN $, $test \gets 1$,    
                \STATE CF = $fsu_{cN,nN,k-1}+ fsu_{cN,nN,k+fs_m+1}$
                \STATE Update change of fragments, $F_c$ as -1,0,1 corresponding to CF= 2,1,0
            \ENDIF
        \ELSE
            \STATE Calculate probabilities $P_u$ for unvisited nodes,$u$
            \IF{$\sum P_u == 0$}
                \STATE Exit  \textcolor{teal}{No feasible Node}
            \ELSE
                \STATE Normalize probabilities: $P \gets P_u/\sum P_u$
                \STATE Choose a next node $nN$ using roulette wheel based on probabilities calculated in (\ref{Probability_Equation_Edge})
                \IF{$fsu_{cN,nN,q}==0, ~ \forall q \in [k:k+fs_m]$}
                    \STATE Set $test \gets 1$
                    \STATE CF = $fsu_{cN,nN,k-1}+ fsu_{cN,nN,k+fs_m+1}$
                    \STATE Update change of fragments, $F_c$ as -1,0,1 corresponding to CF= 2,1,0
                \ELSE
                    \STATE Exit \textcolor{teal}{Continuity issue}
                \ENDIF
            \ENDIF
        \ENDIF
        \STATE Update distance: $dis \gets ant[a].dis + d_{cN,nN}$
        \IF{$nN \neq d$ \textbf{and} $test == 1$ \textbf{and} $dis < d_m$}
            \STATE Update ant[a].Tour, dis, $\Delta F$.
        \ELSE 
            \IF{$nN == d$ \textbf{and} $test == 1$ \textbf{and} $dis < d_m$}
                \STATE Update ant[a].Tour,dis, $\Delta F$.
                \STATE Exit \textcolor{teal}{Successful Traverse}
            \ELSE
                \STATE Exit \textcolor{teal}{Optical reach issue}
            \ENDIF
        \ENDIF
    \ENDFOR
\ENDFOR
\STATE \textbf{return} all Ants
\end{algorithmic}
\end{algorithm}

In the Algorithm 2, the ants are initialized from the source node and are allowed to explore the network (ref: line 1 and 2). The selection of the first intermediate node differs from the rest of the nodes as it involves choosing an auxiliary link from among multiple links connected to the source node with probabilities given in (\ref{Probability_Equation_Aux_Edge}). Using the roulette wheel, we now chose an auxiliary link and subsequently update the ant's information (ref: lines 2 to 11). To efficiently store the information gathered from the ants, we have created a structure named ant[a].X, where $X$ is the variable that contains the information for a specified parameter associated with the ant $a$ (detailed description of which is given in subsection \ref{Info}). In scenarios where auxiliary links are absent at the source, the algorithm is promptly terminated, recognizing the unavailability of a viable path (ref: line 8). Upon identification of the initial intermediate node, the subsequent node is determined using the roulette wheel with probabilities derived from (\ref{Probability_Equation_Edge}) (ref: line 21). Upon the node selection, the contiguity and continuity constraints are verified and the change of fragments are calculated(ref: lines 22 to 29). Thereafter we follow the flow by updating the ant's information and checking the distance constraint. This process continues until the ant reaches its destination or encounters an impasse (ref: lines  18, 27, 37 and 39). To prevent the formation of loops, we set probabilities of the visited nodes as zeroes (ref: line 16). This exploratory procedure is repeated for all the ants.
\subsubsection{Ants Information collected}
\label{Info}
The information stored for each ant in the structure ant[a].X of Algorithm 2 across different parameters are explained below in detail.

\textbf{(fsuindex, mod):} In line 11, we stated that the ant information for the FSU index and the modulation format is updated. This information is collected after the first auxiliary link is found. Since the auxiliary link from the source contains details about the modulation and FSU index, this information is then updated in the \textit{ant[a].fsuindex} and \textit{ant[a].mod}.

\textbf{(Tour, dis, $\Delta F$):} It is noteworthy to mention here that in lines 33 and 36 the information regarding the tour, distance, and fragment changes are updated for an ant. These three parameters are updated after each successful discovery of nextNode (nN). Specifically, ant[a].Tour is updated by appending nN to the existing ant[a]. Tour array and ant[a].dis is updated by adding the link distance from the current node (cN) to nN to the existing ant[a].dis. ant[a].$\Delta F$ is updated by adding $F_c$ to the existing ant[a].$\Delta F$ that is calculated in line number 13 and 25.

\subsection{Calculating fitness function of Ants}
 After the completion of network exploration by all ants, the subsequent step involves evaluating the fitness of those ants that have successfully reached the destination. This evaluation is essential to identify optimal solutions and to offer ants in the next iteration an enhanced opportunity to improve the solution, if possible. The fitness value of each ant $a$ that has reached the destination is determined using the fitness function specified in (\ref{FitnessFunction}) and represented as shown in (\ref{FitnessFunction_of_ant_A}).

\begin{equation}
    ant[a].fitnessFunction = \frac{\Delta F_a}{2LT_a} + FS_{m_a} * LT_a
    \label{FitnessFunction_of_ant_A}
\end{equation}

\noindent where

$\Delta F_a = ant[a].\Delta F$, 

$LT_a = length(ant[a].\textit{Tour})-1 $,

$m_a = ant[a].\textit{mod} $

\noindent All these ants information is returned by Algorithm 2.

\subsection{Updating pheromone concentrations}
The concentration of pheromones in links plays a critical role in the ACO process. Therefore, it is crucial to update the pheromone values with precision following each iteration to enhance solution quality and accelerate the convergence of algorithm. The pheromone update mechanism involves deposition as described in (\ref{Pheromone_update}) and (\ref{Pheromone_update_Aux_link}). Evaporation of pheromone in links is regulated by the evaporation parameter $\sigma$ in (\ref{pheromone_evoporation}) and (\ref{pheromone_evoporation_Aux_link}).

\begin{equation}
    u\tau_{i,j} = u\tau_{i,j} + \frac{1}{ant[a].fitnessFunction}, \forall a
    \label{Pheromone_update}
\end{equation}

\begin{equation}
    u\tau_{s,j,m_l}^k = u\tau_{s,j,m_l}^k + \frac{1}{ant[a].fitnessFunction}, \forall a
    \label{Pheromone_update_Aux_link}
\end{equation}

\begin{equation}
    u\tau_{i,j}= (1-\sigma)u\tau_{i,j} ~\forall~(i,j) \in E.
    \label{pheromone_evoporation}
\end{equation}

\begin{equation}
    u\tau_{s,j,m_l}^k= (1-\sigma)u\tau_{s,j,m_l}^k ~\forall~(i,j) \in E, ~ \forall k \in N,~ \forall m_l \in M.
    \label{pheromone_evoporation_Aux_link}
\end{equation}

\subsection{Exploitation}

As previously mentioned, exploitation is a critical aspect of metaheuristic algorithm as it aims to find improved solution by utilizing previously acquired knowledge. Using the information gathered by ants during their initial exploration, we updated the pheromone values to $u\tau_{i,j}$ and $u\tau_{s,j,m_l}^k$, which we then utilized for further exploitation. The exploitation process follows the same procedure outlined in Algorithm 2, but the probabilities for selecting edges and auxiliary edges are adjusted based on the updated pheromone values. To perform exploitation, we replace $\tau_{i,j}$ with $\tau_{s,j,m_l}^k$ and $u\tau_{i,j}$ with $u\tau_{s,j,m_l}^k$, and then follow Procedure in Algorithm 2. 

\subsection{Termination}
Once the pheromone values are updated, the exploration and exploitation is performed by the ants with their respective pheromone values. This process continues for a predetermined number of iterations, as typically done in conventional ACO, after which the solution is considered coverged. In contrast, our method terminates the ACO process based on various conditions prior to each iteration to quickly identify a solution.
\begin{itemize}
    \item The first condition is verified even before the first iteration: if there is no auxiliary link available from the source, it is evident that a solution cannot be found. The algorithm then terminates immediately, blocking that connection without further calculations.
     \item The second termination condition is assessed after the second iteration and continues through to the final iteration. If 40\% of the ants have identical fitness values, and these values are the lowest among all fitness values, it is assumed that the solution has converged. Consequently, the process is terminated to conserve time and computational resources.

     \item The third termination condition occurs when all iterations are completed, at which point the ACO process ends.

\end{itemize}

\subsection{Final Allocation}

Once the ACO process is terminated, the final allocation of a connection is determined based on the various termination conditions.

\begin{itemize}
    \item  Blocking Scenario: If the ACO process is terminated by the first condition, the requested connection is blocked. Additionally, if the fitness values calculated after termination are not finite, it indicates that no solution is found, and the connection is blocked.

    \item  Allocation Scenario: If the fitness value obtained is finite, either from the second or third termination condition, it suggests that at least one possible solution exists with the minimum fitness value achieved. If the solution is found to be unique (i.e., the route followed and the FSU considered are the same), the connection is allocated using that solution. In case the minimum fitness value corresponds to multiple solutions, the least indexed FSU is selected for the connection allocation. This approach aims to preserve the FSUs at higher indices.
 
\end{itemize}

The complete heuristic algorithm is shown in the below flowchart

\begin{figure}[h]
    \centering
    \includegraphics[width=\columnwidth,keepaspectratio]{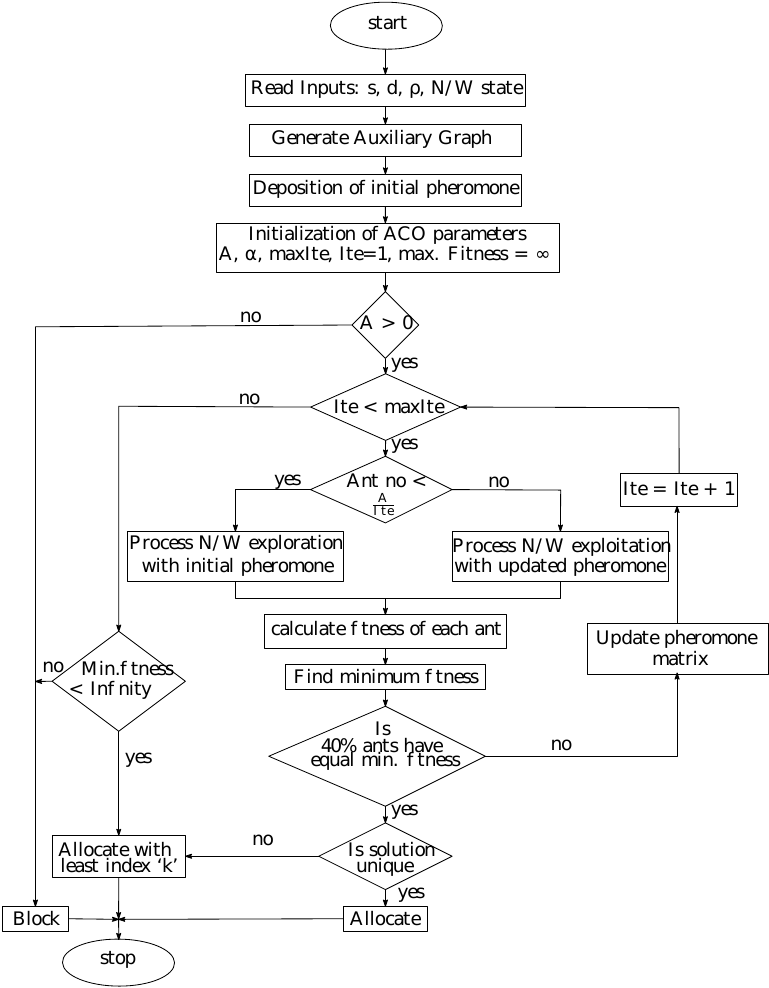}
    \caption{Proposed RSA Algorithm with ACO}
    \label{fig:my_label}
\end{figure}

\subsection{Complexity Analysis}
We assume that there are V nodes, L links, N FSUs, A ants and T iterations. Initially the Auxiliary Graph is generated, followed by the application of the ACO algorithm on top of it. The overall time complexity is found to be the sum of these two individual time complexities. The time complexity for generating the Auxiliary Graph is $O(VN)$, and the complexity for ACO is $O(AT(VN + V^2 + 2L))$. We arrive at this complexity from the following parts of the algorithm: (a) $VN$ is the time complexity involved in picking an auxiliary link connected to the source, (b) $V^2$ is the time complexity for the formation of the path from the rest of the nodes and (c) $L$ for both calculating the fitness function and updating the pheromone values in the path. This procedure is repeated for all ants $A$ and executed for all iterations $T$. Thus, the total complexity is $O(VN + AT(VN + V^2 + L) )$.

However, the complexity of the method described in \cite{comp_nan2022routing} is $O(T(A(V^2+L+A log A))+ kLN)$ where $k$ represents the total number of paths. Further, it may be noted that the usual time complexity of a k-shortest path with the first-fit algorithm is $O(KV(L+VlogV)+ kLN)$.

\ifCLASSOPTIONcaptionsoff
  \newpage
\fi

\section{Results}

\begin{figure*}[h!]
    \centering \subfigure{\includegraphics[width=0.41\textwidth,keepaspectratio]{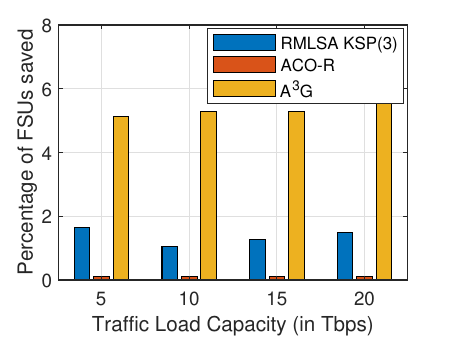}}    \subfigure{\includegraphics[width=0.4\textwidth,keepaspectratio]{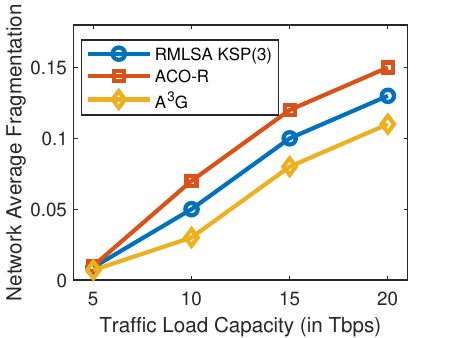}}    
    \caption{Performance analysis in 14-node NSFNET for different load vs.(a) Percentage of FSUs saved, (b) Network Average Fragmentation}
    \label{14_Node_Simulation_Results_L}
\end{figure*}
In this section, we made three distinct analyses and present the results. The first analysis involves a detailed examination with algorithmic parameter `z'. The second analysis provides a comparative assessment between the proposed algorithm and the existing ones keeping the hold time infinite. The third analysis evaluates the performance of algorithm's fitness function under different arrival with finite hold times. We use MATLAB R2023b and $12^{th}$ over Gen Intel(R) Core(TM) i5-1235U processor for all the simulations.
\subsection{Results corresponding to z}
\begin{figure}[h]
    \centering
    \includegraphics[width=1\columnwidth,keepaspectratio]{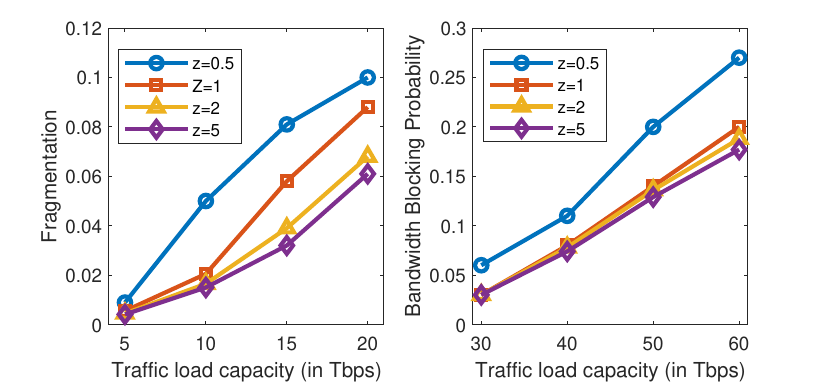}
    \caption{Fragmentaion and Bandwith Blocking Probability with `z'}
    \label{Z_Result}
\end{figure}

In section II, we stated that the total number of ants populated is dynamic and dependent on (a) the network's state and (b) the scaling parameter $z$. We further analyzed the proposed algorithm for different values of $z$ under varying network states. The results corresponding to the bandwidth blocking probability and average network fragmentation are presented in Fig. \ref{Z_Result}. As there is no significant improvement either in bandwidth blocking probability or in the average network fragmentation when the value of $z$ exceeds 2, we fix this value as the scaling parameter for the proposed algorithm.

 The primary termination condition states that the algorithm halts after reaching the maximum number of iterations. Additionally, the second termination condition is dynamic where the algorithm gets halted when $40\%$ of ants have same fitness values. Moreover, we performed the experiment for over 8000 connections in different scenarios, and it was observed that the fitness value does not improve beyond 4 iterations. Therefore, we have fixed the maximum number of iterations at 5 and performed all the simulations accordingly. All these analysis were performed keeping the hold time infinite.

\subsection{Comparative analysis with infinite hold time}

\begin{figure}[h]
    \centering
    \includegraphics[width=1\columnwidth,keepaspectratio]{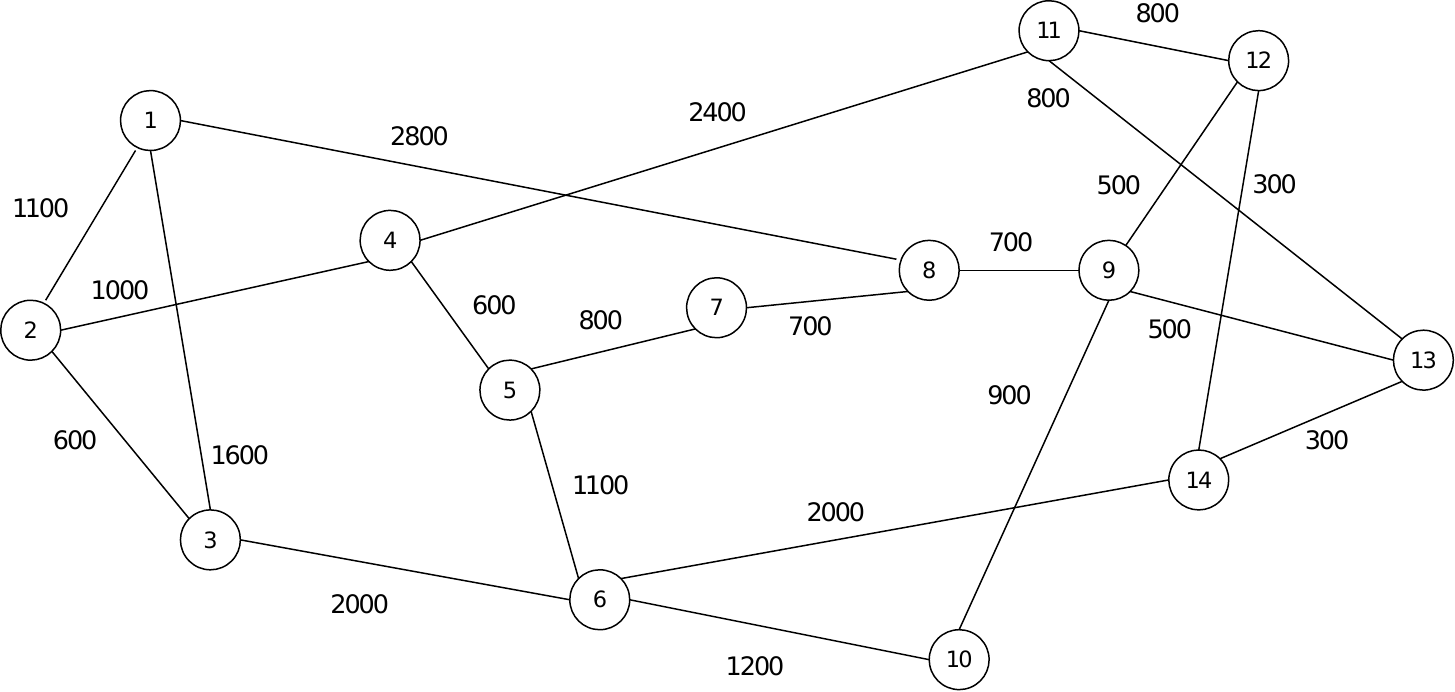}
    \caption{14 Node NSFNET}
    \label{NSFNET}
\end{figure}

\begin{figure*}[t]
  \centering 
    
    \subfigure{\includegraphics[width=0.4\textwidth,keepaspectratio]{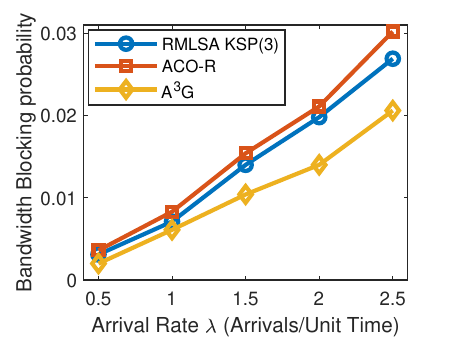}}  
   \subfigure{\includegraphics[width=0.4\textwidth,keepaspectratio]{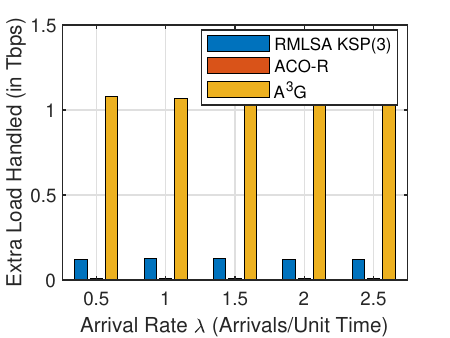}}
    
    \caption{Performance analysis in 14-node NSFNET for different arrival load vs.  (a) Bandwidth Blocking Probability, (b) Extra Load Handled (in Tbps)}
    \label{14_Node_Simulation_Results}
\end{figure*}

In this subsection, a comparative analysis is conducted between the proposed algorithm ($A^3G$), an algorithm (ACO-R) that utilizes ACO solely for routing, and the k-shortest path algorithm with 3 paths ILP formulation (RMLSA KSP(3)) proposed in \cite{comp_nan2022routing}. For this comparative analysis, we have considered the 14-node NSFNET as shown in Fig. \ref{NSFNET} with all distances provided in kilometers. We consider the optical reach as 3600 km for BPSK, 2400 km for QPSK, 1200 km for 8-QAM, and 600 km for 16-QAM \cite{comp_nan2022routing}. We also consider the evaporation parameter $\sigma$ to be $0.5$. Moreover, while performing simulations we assume that both source and destination nodes were selected from a uniform distribution. The arrival process is modeled as a Poisson process, where the inter-arrival times follow an exponential distribution. The arrival rate is denoted by $\lambda$. The holding times, represented as $t_h$, also follow an exponential distribution with the parameter $\mu$ (service rate), which is the inverse of the holding time. Data rates ranging from 50 Gbps to 500 Gbps were also drawn from a uniform distribution for each specific source-destination pair. The total number of slots considered for each link in the network is 320.

We generated a comprehensive set of simulation results by averaging over 15 distinct input sets mentioned above. We observed that none of the connections for all formulations are blocked up to an aggregate data rate of 20 Tbps with holding time $t_h$ to be very high. Up to this threshold, calculations were performed for (a) the average number of FSUs saved and (b) the network average fragmentation (NAF). The mathematical expression of NAF is taken from \cite{comp_nan2022routing} and is calculated as (\ref{NAF}).

\begin{equation}
   NAF= \frac{1}{L}\sum_{\forall (i,j) \in E}^{} (1-\frac{\textit{Largest Fragment Block}_{i,j}}{\textit{Total Vacant slots}_{i,j}})
   \label{NAF}
\end{equation}

In the context of spectrum utilization, the proposed formulation consistently achieves close to 5.4\% reduction in FSUs and a minimum of 14\% reduction in NAF compared to KSPF(3) and ACO-R under similar simulation scenarios, as shown in Fig. \ref{14_Node_Simulation_Results_L}(a) and \ref{14_Node_Simulation_Results_L}(b). This significant improvement is due to the network being explored based on requirements without a predefined path and the fitness function that effectively minimizes fragments. Thus, it results in better NAF while maintaining optimal spectrum utilization.

\subsection{Comparative analysis with finite hold time}

The network's steady state analysis is key to understanding dynamic traffic at varying input arrival rates when the holding time is finite. The arrival process is modeled as a Poisson process. The arrival rate is $\lambda$ and the holding time is $t_h$. In steady state, due to stationarity, the measurable network parameters are statistically consistent across different time samples for a given arrival rate, with the holding time kept unchanged. So, we analyze the network at different arrival rates, ranging from 0.5 arrivals/unitTime to 2.5 arrivals/unitTime, with a holding time of 2 unitTimes. We present results related to Bandwidth Blocking Probability and the average extra load handled, considering various models under the same arrival rates.

The Bandwidth Blocking Probability (BBP) at any given moment is expressed in (\ref{BBP}). After the network reaches a steady state, the calculated BBP is averaged over a specified time window to determine the average BBP for a given arrival rate. It is observed that the BBP of the proposed model is approximately 13 percent lower than that of other models at an arrival rate of 1 arrival/unitTime. This difference increases to approximately 34 percent at an arrival rate of 2.5 arrivals/unitTime as shown in Fig.\ref{14_Node_Simulation_Results}.(a).    

\begin{equation}
    BBP = \frac{\text{Requested Data Rate Blocked}}{\text{Total Load in the network}}
    \label{BBP}
\end{equation}

The average extra load that the network handles for any given formulation is the difference between the average load managed by that formulation and the smallest average load among all formulations. Thus, the formulation with the smallest average load will result in zero extra load. Fig.\ref{14_Node_Simulation_Results}.(b) illustrates that the additional load managed by $A^3G$ at various arrival rates is approximately 1.1 Tbps. This extra load handled and the reduction in BBP is a reflection of the resource savings and the reduced network average fragmentation, as depicted in Fig. \ref{14_Node_Simulation_Results_L}.(a) and Fig. \ref{14_Node_Simulation_Results_L}.(b) respectively.

\section{Conclusion}
In this paper, we address the RMLSA problem for dynamic scenarios by jointly solving it using a metaheuristic algorithm inspired by ACO. By incorporating the constraints of RSA in EONs, we modified conventional ACO into a constraint-based ACO. This process is fully adaptive at all stages, depending on the inputs and the state of the network. We have selected a fitness function designed to optimally utilize the spectrum while maintaining lower average network fragmentation.

Our results demonstrate a consistent $5.4 \%$ savings in spectrum utilization, a $14 \%$ lesser network average fragmentation compared to recent RSA solutions. The enhanced network average fragmentation and effective spectrum utilization provide a significant advantage, enabling the handling of an additional 1.1 Tbps of load across varying arrival rates. This improvement also results in a reduction of the BBP by $13\%$ at low arrival rates and $34\%$ at high arrival rates. This improved network average fragmentation and the effective spectrum utilization  are achieved through the joint approach driven by the nature of the fitness function. The improvement in the results supports the conclusion that solving the RMLSA problem jointly yields better performance than relying on the predetermination of a fixed number of paths. This joint approach, when coupled with enhanced adaptation at every stage and strategic graph augmentation, produces superior results and achieves quicker convergence.

\bibliographystyle{IEEEtran}
\bibliography{ref}

\end{document}